\documentstyle[preprint,aps,epsfig,12pt,tighten,amssymb]{revtex}
\def\DESepsf(#1 width #2){\epsfxsize=#2 \epsfbox{#1}}
\begin{document}
\draft
\renewcommand{\thepage}{-- \arabic{page} --}
\preprint{KUHD/00-8}
\title{
On a possible manifestation of $f_1$ trajectory 
in $J/\psi$ photoproduction}
\vfill
\author{
{\ Dayeeta Roy}$^{1)}$
\thanks{Email address: roy@radix.h.kobe-u.ac.jp},
{\ Toshiyuki Morii}$^{1),2)}$
\thanks{Email address: morii@kobe-u.ac.jp}
and
{\ Alexander I. Titov}$^{3)}$
\thanks{Email address: atitov@thsun1.jinr.ru} \\
\vspace*{0.5cm}
}
\address{
1) Graduate School of Science and Technology, Kobe University\\
    Nada, Kobe 657-8501, JAPAN\\
2) Faculty of Human Development, Kobe University \\
   Nada, Kobe 657-8501, JAPAN\\
3) Bogolyubov Laboratory of Theoretical Physics, JINR\\ 
141980 Dubna, Russia \\
\vspace*{0.2cm}
}
\date{\today}
%
%
\vfill
\maketitle
\thispagestyle{empty}
\begin{abstract}
We analyze a possible manifestation of $f_1$-trajectory 
in elastic $J/\psi$ photoproduction 
at high energy and large momentum transfer. Inspite of  
the small contribution of $f_1$-trajectory in total cross sections, 
it becomes significant in various spin observables. 
In particular, we show that the crucial test 
for $f_1$-exchange  can be made by measuring the 
single beam- and double parity- and beam-target
asymmetries at large momentum transfers, 
where a strong deviation from 
the exchange of conventional Pomerons is expected. 
This  effect is caused by the interference 
of natural (Pomeron) and  unnatural ($f_1$) parity exchange parts 
of amplitude in the region where their contributions become 
comparable to each other and might be interesting to observe in forthcoming experiments, if feasible.
\\
\end{abstract}

\vskip2pc
\setcounter{page}{1}

\pagestyle{plain}
\indent
Hadronic diffraction has regained popularity 
in the recent years due to many new interesting experiments 
at HERA and TEVATRON. One of them  is the vector 
meson photoproduction at HERA. 
This channel is exciting because here 
one can study any of the known vector mesons from the 
lightest $\rho$ to heavier $J/\psi$ and $\Upsilon$ and  
hope to see the transition from the ``soft'' to the ``hard'' 
regime. The most popular models for description of the diffractive 
processes at high energy are based on the
Regge theory~\cite{REG} where the corresponding 
multiparticle exchanges with definite quantum numbers 
are expressed by the effective exchange of one Regge 
pole whose propagator is given by $({s}/{s_0})^{\alpha(t)}$. 
The Regge trajectory $\alpha(t)$ has a simple linear 
form $\alpha(t)=\alpha(0)+\alpha't$ and the main contribution to the 
elastic forward photoproduction of light vector mesons comes 
from the Pomeron (so-called ``soft'' Pomeron) trajectory 
with intercept 1.08 and slope of 0.25 GeV$^{-2}$~\cite{DL}.
This leads to rather  weak energy dependence 
of the cross section gradually 
becoming almost constant at very high energies. However, the Pomeron, though
amply used to successfully analyze the high energy total hadron cross
sections, still remains mysterious in its partonic content, 
hence its manifestation in QCD also remains unknown. 

Recent studies of elastic vector meson photoproduction data shows that 
the $J/\psi$ cross-section shows a steep rise with the energy
$W$~\cite{DAT}. This behavior is quite different from that of other 
light vector mesons like $\rho,\omega$ and $\phi$ which are characterized
by a weak dependence of their cross section on energy
[$\sigma \propto W^{0.22}$], while for  $J/\psi$ it is
parameterized as $W^{0.8}$. This new phenomenon 
drew a lot of attention in this field leading to several 
models being proposed to explain the data~\cite{DLH,RYSK,BROD,RYSK2}.
However, this subject still remains controversial. Donnachie and
Landshoff~\cite{DLH} claimed that perturbative QCD does not work in
this energy region and suggested the  use of a small admixture of 
the second ``hard'' Reggeon  exchange whose trajectory is given by
$\alpha_h(t)= 1.418 + 0.1\,t$. 
The conventional mesonic trajectories arising from the fits of Donnachie
and Landshoff (for instance, $f_2$-trajectory~\cite{DL,KMRV}) 
contribute to light vector meson productions at relatively low
energies. Most of them have a slope of 0.9 GeV$^{-2}$ and 
an intercept $\alpha(0) \simeq 0.5$. 
These trajectories with positive and negative 
signature are important in the near-threshold region though 
their contribution decreases with energy. 
But for description of the $J/\psi$ photoproduction for all energy 
regions, one must take into account all of the above trajectories: Pomeron, 
hard Reggeon and mesonic (supersoft) trajectories~\cite{US}.

The  Pomeron and ``hard'' 
Reggeon which are dominant at high energy
on the phenomenological level
are described by the effective  ``$C=+1$  photon''  
exchange  which leads to similar spin observables for all components
typical for the amplitude with  $t$ - channel natural parity exchange.
However, so far the study of the unnatural parity
exchange trajectories at high energy has not been done
widely. Recently, an interesting analysis for testing the new
$f_1$-trajectory associated with the axial vector meson exchange was
done by Kochelev et al.~\cite{KOCH}, in which the slope of the
trajectory was taken to be almost zero. It is very interesting to test
the effect of this trajectory even for heavy vector meson $J/\psi$
photoproduction. Since its slope is close to zero, it may be active in
the high energy and large momentum transfer region.

In this work, we are concerned with the large momentum transfer region, where 
the conventional soft Pomeron, the hard Reggeon and $f_1$ trajectory
contribute. We show that the spin observables are sensitive to
$f_1$-trajectory and will help to differentiate the different components
of the amplitude in this region. Let us start with the definition of the 
kinematical variables for the $\gamma p \rightarrow J/\psi p$ reaction
using standard notation. The four momenta of the incoming photon,
outgoing $J/\psi$, initial (target) and final (recoil) proton are
represented by $k,q,p$ and $p^{\prime}$, respectively. Hereafter,
$\theta$ denotes the $J/\psi$ production angle in c.m.s. and $s \equiv
W^2 = (p + k)^2, M_N$ is the nucleon mass, $M_V$ is the $J/\psi$ mass
and $m_{f_1}$ is the $f_1$-meson mass. We use the convention of Bjorken and
Drell to define the $\gamma$ matrices and Dirac spinors.

For simplicity we call the ``soft'' Pomeron and the  ``hard'' Reggeon
trajectories [Donnachie-Landshoff(DL)] as the  Pomeron
part~\cite{DL,DLH}. The corresponding invariant amplitude in the standard notation reads,
\begin{eqnarray}
T_{fi}^{P}=i\sum_{i=s,h}\bar{u}_{m_{f}}({p}')
M_{R_{i}}\epsilon^\ast_{V\mu}(\lambda_f) 
( (\gamma\cdot k)g^{\mu\nu} - \gamma^\nu k^\mu)
\epsilon_{\gamma\nu}(\lambda_i)u_{m_{i}}(p), 
\label{TP}
\end{eqnarray}
The factor $M_{R_{i}}$  is given by the conventional 
Regge pole amplitude,
\begin{eqnarray}
M_{R_{i}}=
12\,a_i\,s_0 \,\beta_0^2
\left(\frac{3\Gamma_{V\to e^+e^-}}{\alpha_{\rm em}M_V}\right)^{1/2} 
\,F_N(t)\,F_V(t) e^{-i\frac{\pi}{2}(\alpha_i(t)-1)} 
\left(\frac{s}{s_0}\right)^{\alpha_i(t)},
\label{ampl}
\end{eqnarray}
where $i=s,h$ for soft Pomeron and hard Reggeon, 
respectively. $V \equiv J/\psi$; 
$s_0=4$ GeV$^2$, $\beta_0=1.64$ GeV$^{-1}$; $a_s=1$, $a_h=0.05$; 
$\alpha_{\rm em}=1/137$, $\Gamma_{J/\psi\to e^+e^-}$ = 5.26 KeV~\cite{PDG}.
The trajectories of these Pomerons are given by,
\begin{eqnarray}
\alpha_s(t)=1.08 +0.25\,t,\nonumber\\ 
\alpha_h(t)=1.418 + 0.1\,t.
\end{eqnarray}
$F_N$ is the isoscalar electromagnetic form factor of the nucleon and
$F_V$ is the form factor for the vector-meson-photon-Pomeron 
coupling~\cite{FF,LAGET}
\begin{eqnarray}
F_{N}(t) &=& \frac{(4M_N^2-2.8t)}{(4M_N^2-t)(1-{t}/{0.7})^2},
\label{FN}\\
F_{V}(t)&=&\frac{M_V^2}{(M_V^2-t)^2}\,\frac{\mu^2}{2\mu^2 + M_V^2-t},
\label{FV}
\end{eqnarray}
with $\mu=1.1$ GeV$^2$.

For the $f_1$-trajectory, we follow  the consideration of 
Ref.~\cite{KOCH} where  it is assumed that the intercept 
of this trajectory is equal to 1, in agreement with the low $x$ behavior
of the spin-dependent deep  inelastic structure 
function $g_1(x,Q^2)\sim 1/x^\alpha$ with
$\alpha=0.9\pm 0.2$. Moreover it is assumed that the slope 
of the trajectory is close to zero, which is in agreement with
large $t$ behavior of elastic $pp$ scattering. This means that
the excited  meson states of this trajectory have very large masses
and the total amplitude with unnatural parity is saturated 
by the lowest $f_1(1285)$ meson exchange which may be expressed in terms of the one boson 
exchange amplitude. The effective vertices of the axial-vector $f_1$-meson
interaction with the nucleon  and the $f_1$-photon-vector-meson
interaction reads,
\begin{eqnarray}
V_{f_1NN}&=&ig_{f_1NN}\bar u_f\gamma_\mu\gamma^5\bar u_i\,
\xi^\mu,\\
V_{f_1V\gamma}&=& g_{f_1V\gamma}\epsilon_{\mu\nu\alpha\beta}
\xi^{\beta}\epsilon_1^{\nu}\epsilon_2^{\alpha}M_V^2k^{\mu}
\label{Vf},
\end{eqnarray}
where $\xi$ and $\epsilon$ are the polarization vectors of 
the axial vector and vector mesons, respectively, 
and 
$g_{f_1NN}=2.5$ is fixed from the proton spin analysis.
The coupling constant $g_{f_1V\gamma}$ is extracted 
from the radiative decay $J/\psi \rightarrow f_1\gamma$,
\begin{equation}
\Gamma_{V\rightarrow f_1\gamma}
=g_{f_1V\gamma}^2
\frac{(M_V^2 - M_{f_1}^2)^3 (M_V^2 + M_{f_1}^2)}{96\pi M_V M_{f_1}^2}.
\end{equation}
Taking $\Gamma_{J/\psi\rightarrow f_1\gamma}=56.5$ eV~\cite{PDG},
one gets $|g_{f_1 J/\psi\gamma }| = 1.245 \times 10^{-4}$ GeV$^{-2}$.\\
Using the above vertex functions and coupling constants, 
the corresponding matrix element of the vector meson production 
can be written as,
\begin{eqnarray}
T_{fi}
= ig_{f_1V\gamma}g_{f_1NN}
F_{f_1NN}(t)F_{f_1V\gamma}(t)
\frac{m_V^2}{t - m_{f_1}^2}
\epsilon_{\mu\nu\alpha\beta}q^{\mu}\epsilon_V^{*\nu}(\lambda_V)
\epsilon_{\gamma}^{\alpha}(\lambda_{\gamma})\nonumber\\
\times \left( g^{\beta\delta} 
- \frac{(p-p^{\prime})^{\beta}(p-p^{\prime})^{\delta}}
{m_{f_1}^2}\right) \bar{u}_{m_f}(p^{\prime})\gamma_{\delta}
\gamma_5 u_{m_i}(p),
\label{Tf}
\end{eqnarray}
where $m_{f_1} = 1.285$ GeV.
For the $f_1NN$ vertex, we use the flavor singlet axial vector 
form factor which is fixed from fitting to elastic
$pp$ scattering at high $\sqrt s$ and large $|t|$~\cite{KOCH}.
(See also~\cite{SKY}.)
\begin{eqnarray}
F_{f_1NN}(t) = 1/(1 - t/m_{f_1}^2)^2.
\label{FfNN}
\end{eqnarray}
The form factor of the $f_1V\gamma$ vertex is taken as
\begin{equation}
F_{f_1V\gamma}(t) = \frac{\Lambda_V^2 - m_{f_1}^2}{\Lambda_V^2 - t}
\label{FfVg}
\end{equation}
where $\Lambda_V$  for $J/\psi$ is estimated to be
$\Lambda_{J/\psi}=4.2$ GeV by extrapolating 
the $\Lambda_V$ found for $\rho$ and $\phi$ mesons to the one for
the heavy $J/\psi$ meson assuming a linear dependence. 

Note that the $t$-dependence of Eq.(\ref{Tf}) 
is flatter than that of Eq.(\ref{TP}). One reason for this is that the
conventional Pomerons have finite trajectory slopes ($\alpha_i>0$)
whereas the slope of $f_1$ trajectory is almost zero. 
A slight difference in
$t$-dependence of the corresponding form factors $F_NF_V$ and $F_{f_1NN}
F_{f_1V\gamma}$ also play a part in determining the overall $t$-dependence.
  
\begin{figure}[h]
  \begin{center}
  \epsfxsize=6cm \epsfbox{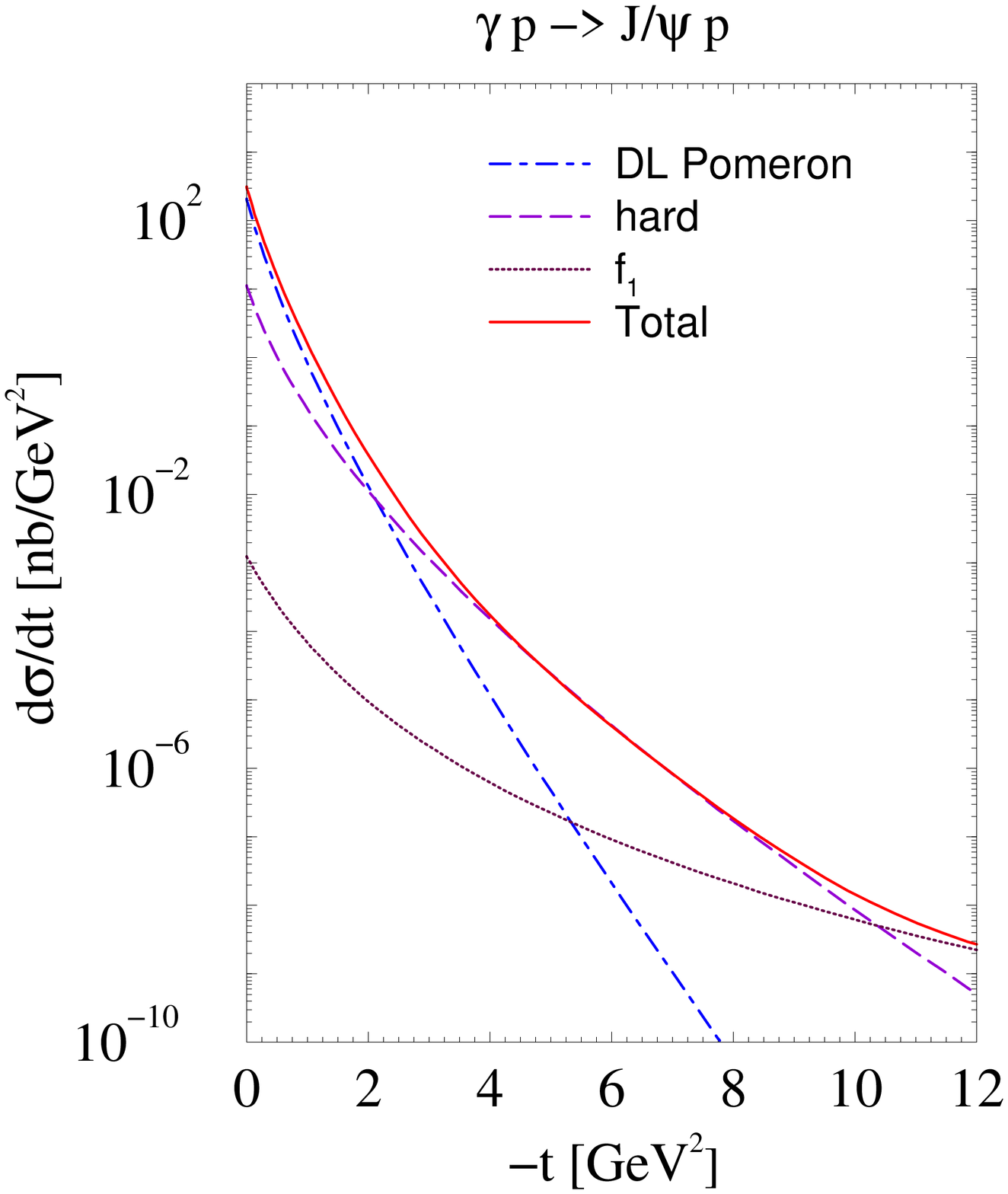}\qquad\qquad
  \epsfxsize=6cm \epsfbox{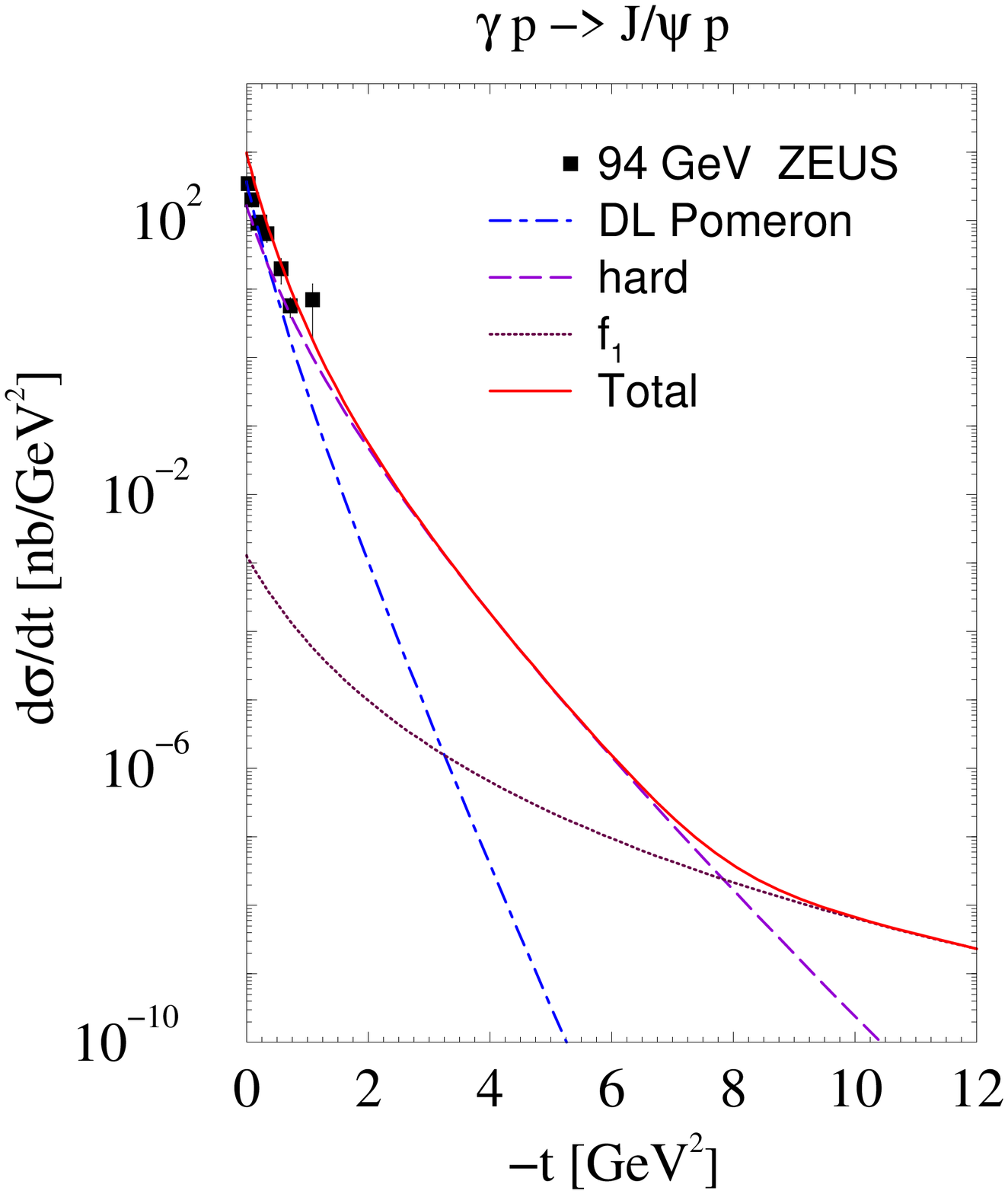}
\end{center}
\vspace*{0.2cm}
\caption
{The differential cross section as 
a function of $-t$ 
for $W=20$ GeV (left panel) and 94 GeV (right panel). 
The dot-dashed and long-dashed lines are the soft Pomeron and hard
 Reggeon exchanges, 
the dotted line is the $f_1$ 
trajectory and the solid line represents the sum of all
channels. 
The data points at the  right panel 
for $W=94$ GeV are taken from~\protect\cite{ZEUS94}.} 
\label{fig:CS}
\end{figure}

Fig.~{\ref{fig:CS}} shows the differential cross sections of $J/\psi$
photoproduction for $W=20$ and $94$ GeV. One can see 
that although the $f_1$-contribution is not conspicuously seen  
at forward production angles (and in total cross sections), 
it becomes comparable with (and dominant over) 
the conventional Pomerons 
in the differential cross sections at large $|t|$. Since its
contribution does not qualitatively change the shape of the cross section,
it is difficult to extract the
$f_1$-exchange from the cross section alone. Therefore, we should study the 
polarization observables for its possible manifestation.

\begin{figure}[h]
\begin{center}
  \epsfxsize=8cm\epsfbox{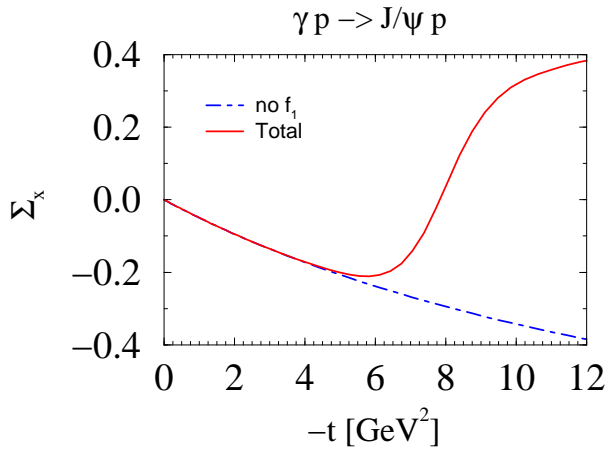} 
\end{center}
\vspace*{0.2cm}
\caption{The beam asymmetry  $\Sigma_x$
as function of $-t$
at $W = 94$ GeV. 
The solid and dot-dashed lines are calculation with and without 
$f_1$ trajectory, respectively.
}
\label{fig:Sigma_x}
\end{figure}

\begin{figure}[h]
\begin{center}
  \epsfxsize=8cm  \epsfbox{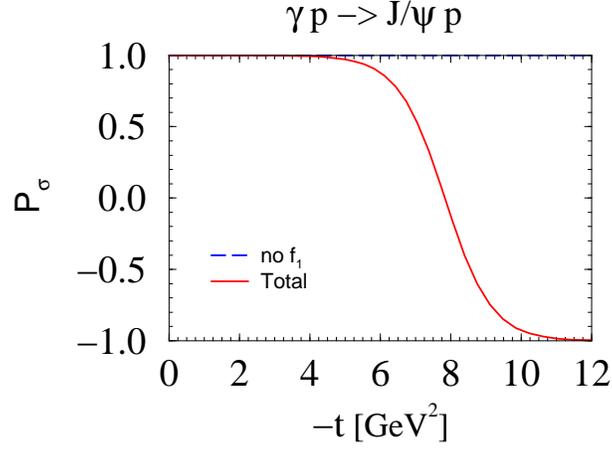} 
\end{center}
\vspace*{0.2cm}
\caption{The parity asymmetry as function of $-t$.
Notation is the same as in Fig.~{\ref{fig:Sigma_x}}.
}
\label{fig:P_s}
\end{figure}

Our analysis of single spin observables shows that the one which is most
sensitive to the $f_1$-trajectory is the beam asymmetry~\cite{TOYM}
\begin{eqnarray}
\Sigma_x\equiv\frac{{\rm Tr}[T\sigma^x_\gamma T^\dag]}
{{\rm Tr}[T\, T^\dag]}
=
\frac{\sigma_{\bot} - \sigma_{\|}}
{\sigma_{\bot} + \sigma_{\|}},
\end{eqnarray}
where $\|\,(\bot)$ corresponds to a photon linearly polarized beam along 
(perpendicular) to the $J/\psi$ production plane. 
The prediction for $\Sigma_x$ is shown in Fig.{\ref{fig:Sigma_x}}.
All the calculations shown below are done for $W=94$ GeV. 
The results shown are for both with (solid curve) and without (dot-dashed
curve) $f_1$ contribution. The pure Pomeron part results in monotonic
decreasing of  $\Sigma_x$ from 0 to some negative value depending on
$t$. Solving Eq.(12) using Eq.(9), for pure $f_1$ channel at high energy and
small vector meson production angle $\theta$, we get, 
\begin{eqnarray}
\Sigma_x^{f_1}
=\frac{s\sin^2{\frac{\theta}{2}}}{s\sin^2{\frac{\theta}{2}} +   M_V^2}
\,>0,
\end{eqnarray}
which results in  strong increase of  $\Sigma_x$ up to
the large positive values exhibiting non-monotonic $t$-dependence.
To further facilitate the experimental test of our predictions, we have
also investigated two other spin observables. 
The first observable is the parity asymmetry defined as~\cite{POL}
\begin{eqnarray}
P_{\sigma} 
\equiv \frac{\sigma^N - \sigma^U}{\sigma^N + \sigma^U} 
= 2\rho_{1-1}^1 - \rho_{00}^1,
\end{eqnarray}
where  $\sigma^N$ and $\sigma^U$ are the cross sections due to the natural
and unnatural parity exchanges and $\rho^i_{\lambda\lambda'}$ are 
the vector meson spin density matrices.
In the region  where the natural parity exchange Pomeron part 
is dominant, 
one expects $P_\sigma=1$. Thus, any deviation from this
value will be due to the unnatural parity $f_1$-exchange. 
Fig.~{\ref{fig:P_s}} shows the corresponding prediction. One can see
strong $t$ dependence of $P_\sigma$ which varies from 
1 to $-1$ in a relatively small interval of $t$.  

\begin{figure}[h]
\begin{center}
    \epsfxsize=8cm \epsfbox{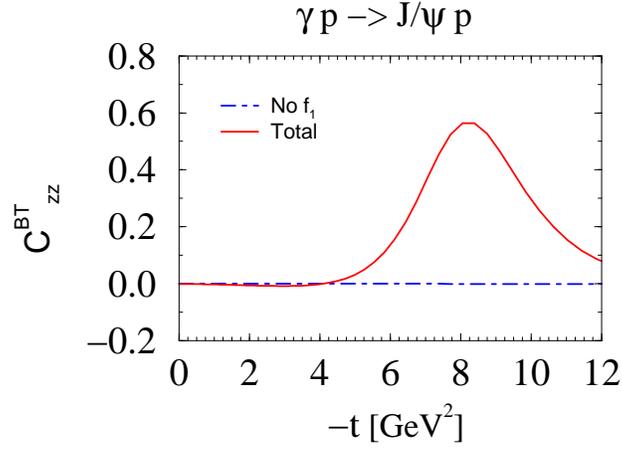}
\end{center}
\vspace*{0.2cm}
\caption {The beam target asymmetry as function on $-t$.
Notation is the same as in Fig.~{\ref{fig:Sigma_x}}.
}
\label{fig:C_BT}
\end{figure}

The second one is the double beam-target asymmetry which 
is defined as~\cite{TOYM},
\begin{equation}
C_{BT} 
\equiv-A_{LL}= \frac{d\sigma (\rightleftarrows )
- d\sigma (\rightrightarrows)}
{d\sigma (\rightleftarrows) + d\sigma (\rightrightarrows)},
\end{equation}
where the arrows denote the relative orientations 
of the proton and photon helicity, respectively. For both pure 
natural and unnatural parity exchange amplitudes, $C_{BT}$ is
exactly zero. The deviation from this value is due to the interference
between natural (Pomeron) and unnatural ($f_1$-exchange) parts 
of the total amplitude.
The corresponding calculation is shown in Fig.~{\ref{fig:C_BT}}.
It can be seen clearly that at some region of $t$, the beam-target asymmetry
deviates from zero considerably. This deviation is proportional to 
${\rm Re}(T_PT_{f_1})$ and can be used to disentangle and identify the
relative strengths of the two parts of the amplitude in the data.

In summary, we have analyzed the possible manifestation of the $f_1$
trajectory in $J/\psi$ elastic photo-production at high energy and large 
momentum transfers. The trajectory parameters are taken from the
prediction of Ref.~\cite{KOCH} and strength $g_{Vf_1\gamma}$ is 
estimated using~\cite{PDG}. It is found that 
the  $f_1$ exchange  contribution can significantly influence 
the differential cross section at large $|t|$. We have presented 
predictions showing the $f_1$ effect on several spin observables.
In particular, we have shown notable effects in  
single beam and double beam-target and parity asymmetries. It should be
noted that our prediction is based on the assumption of the zero slope
of the $f_1$ trajectory discussed in Ref.~\cite{KOCH}, in which two
physical grounds for almost zero slope of the $f_1$ trajectory were
given as (i) there is a nonvanishing contribution to the elastic $pp$
scattering at high energy due to such $f_1$ trajectory and (ii)
$f_1$-meson exchange is deeply related to the axial anomaly which is
associated with the instanton fluctuations whose space scale is much
smaller than the scale of confinement. On the contrary, if the slope of
$f_1$ trajectory is
finite, we expect strong decreasing of its contribution both in
$s$-dependence of the $\sigma_{tot}, d\sigma/dt(t=0)$ and in
$t$-dependence of $d\sigma/dt(t=0)$ at large $s$ and in consequence, any 
deviation from the standard Pomeron model for all the polarization
observables discussed in our paper would disappear.
Experimental test of our predictions will be a useful step
towards understanding the structure of the vector meson
photoproduction amplitude at large $|t|$. To test the effect of
the $f_1$ trajectory contribution alone, it might be advantageous to 
use the other light vector meson $(\rho,\phi)$ photoproductions
discussed in Ref.~\cite{KOCH}. However, if non-trivial interference
effects in spin variables predicted in this work could be observed in
experiment, it will not only be just a supplementary support for the
$f_1$ trajectory but also a good test of the hard Pomeron contribution
as well, because the hard Pomeron is important only in the $J/\psi$ photoproductions.  

We should emphasize that the present investigation is 
the first step from the point of view of a dynamical 
treatment of the problem, as, for example, has been done for effective 
two-gluon model of Pomeron~\cite{DAT2}.
Here, we do not know the sign of $g_{Vf_1\gamma}$ vertex, and have
predicted only the absolute value of beam-target asymmetry. 
Moreover, it would be desired to generate the the corresponding form 
factor independently.

Finally, the experimental facilities like SPring-8, RHIC and HERA have 
plans to work more on polarized vector meson production. Our aim here is 
to find a clear non-trivial effect for the manifestation of the $f_1$
trajectory which reflects some new physics. Without the data on the
experimental/accelerator resources which are needed to perform this
measurement, it is rather difficult to give a detailed and complete
analysis on the feasibility at present. We can say that very high
luminosity, hopefully larger than $10^{32}cm^2 sec^{-1}$, might be
necessary to actually test our prediction since the expected cross
section at large $|t|$ regions where the predicted effect becomes
evident, is quite small.
We hope that such high luminosity could be realized in the forthcoming
experiments so that we can have some good results to support this trajectory 
and also to establish its manifestation in QCD.\\

{\large\bf Acknowledgment}\\

We gratefully acknowledge fruitful discussion with N.I. Kochelev and
Y. Oh. One of us (T.M) would like to thank the Grant-in-Aid 
for Scientific
Research, Ministry of Education, Science and Culture, Japan
(No. 11694081), for financial support.



\end{document}